\documentclass[superscriptaddress, reprint, aps, showpacs, prl]{revtex4-2}

\usepackage{graphicx}
\usepackage{color}
\usepackage{times}
\usepackage{amsmath}
\usepackage[utf8]{inputenc}
\usepackage[colorlinks=true]{hyperref}
\hypersetup{linkcolor=blue,urlcolor=blue,citecolor=blue}

\newcommand{\ieap}{Institut f\"ur Experimentelle und Angewandte Physik, Christian-Albrechts-Universit\"at zu Kiel, 24098 Kiel, Germany}
\newcommand{\syn}{Synopsys Denmark, Fruebjergvej 3, Postbox 4, DK-2100 Copenhagen, Denmark}

\begin{document} 
\title{Resonance-enhanced vibrational spectroscopy of molecules on a superconductor} 

\author{Jan Homberg} \affiliation{\ieap}
\author{Alexander Weismann} \email{weismann@physik.uni-kiel.de}\affiliation{\ieap} 
\author{Troels Markussen} \affiliation{\syn}
\author{Richard Berndt} \email{berndt@physik.uni-kiel.de}\affiliation{\ieap} 

\begin{abstract}
Molecular vibrational spectroscopy with the scanning tunneling microscope is feasible but usually detects few vibrational modes.
We harness sharp Yu-Shiba-Rusinov (YSR) states observed from molecules on a superconductor to significantly enhance the vibrational signal.
From a lead phthalocyanine molecule 46 vibrational peaks are resolved enabling a comparison with calculated modes.
The energy resolution is improved beyond the thermal broadening limit and shifts induced by neighbor molecules or the position of the microscope tip are determined.
Vice versa, spectra of vibrational modes are used to measure the effect of an electrical field on the energy of YSR states.
The method may help to further probe the interaction of molecules with their environment and to better understand selection rules for vibrational excitations.
\end{abstract}

%	07.79.Fc 	Scanning Tunneling Microscopes
%	74.55.+v 	STM single particle tunneling (superconductivity)
% 74.78.Na	superconducting Nanostructures
% 33.20.Tp  molecular vibrational levels
\maketitle 

The vibrational modes of a molecule are sensitive probes of molecular bonds and their response to external parameters. 
They may serve as fingerprints for identification and as indicators of geometric changes.
Molecular vibrations can also help elucidating processes at surfaces.
In 1998, the unequivocal observation of vibrational features in data recorded with the scanning tunneling microscope (STM) was reported \cite{stip}.
Like in the case of molecules in metal-oxide-metal junctions \cite{jako}, the tunneling current $I$ is modified when inelastic processes like molecular vibrations are excited.
Unfortunately, the net effect on the current is usually on the order of few percent at best because the additional inelastic current is partially compensated by a reduction of the elastic current \cite{nico}.
Consequently, only few molecular vibrations lead to observable signals, which typically are best discerned in $dI^2/dV^2$, the second derivative of $I$ with respect to the sample voltage $V$.
A notable exception are data from C$_{60}$ adsorbed to superconducting lead, where 9 modes were observed \cite{kath}.
Vibronic transitions may also lead to larger changes of the tunneling conductance but detailed spectra of various modes have not been reported \cite{dopp, mati, schw}.
In addition, vibrational features may be resolved in single-molecule fluorescence spectroscopy with the STM \cite{qiu, dong}.

\begin{figure}[!h]
\begin{center} \includegraphics[width=0.7\columnwidth]{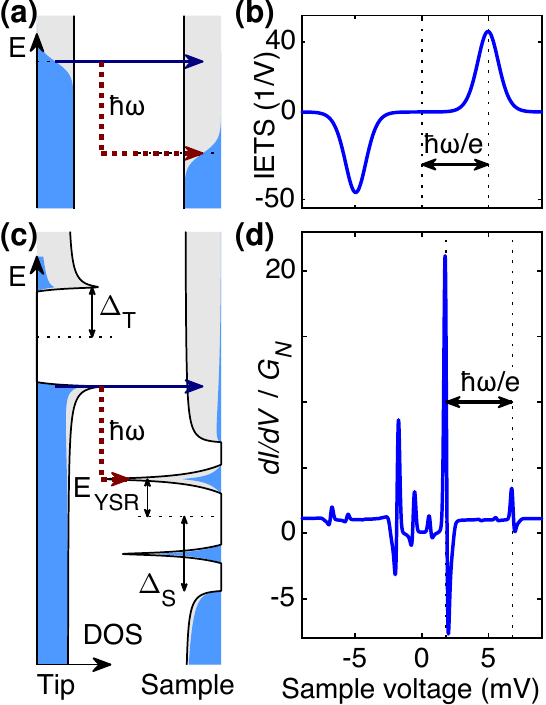} \end{center}
\caption{(color online) (a) Schematic energy diagram of elastic (blue solid) and inelastic (red dashed) tunneling processes between normal conducting tip and sample electrodes with energy independent density of states (DOS).
Both Fermi distributions (occupied states shaded in blue) are broadened by temperature.
The sample voltage $V$ is chosen to match the energy of a vibrational excitation, $eV=\hbar\omega$.
(b) Second derivative $d^2I/dV^2$ of the current $I$ scaled with $dI/dV$ for the scenario in (a) (parameters $\eta = 10\%$, $\hbar \omega = 5$\,meV, $T=4.2$\,K).
%shortened prev sentence
The inelastic process gives rise to a broad peak/dip at $\pm \hbar\omega /e$.
(c) Related diagram for a superconducting tip (energy gap $2 \Delta_\mathrm{T}$) and sample ($2 \Delta_\mathrm{S}$).
A paramagnetic impurity induces sharp YSR states, separated by $\pm E_\mathrm{YSR}$ from the Fermi level $E_F=0$ of the sample (dotted line). 
$V$ is chosen to satisfy $eV=\hbar \omega + E_\mathrm{YSR} + \Delta_\mathrm{T}$.
(d) Calculated spectrum of $dI/dV$ scaled by the conductance $G_N$ outside the gap [$\eta$, $\hbar \omega$, and $T$ as in (b)].
The YSR states lead to two predominant asymmetric peaks with different amplitudes at $V = \pm (E_\mathrm{YSR} + \Delta_\mathrm{T})/e$.
Tunneling of thermally excited electrons ({\em e.\,g.}\ from the upper coherence peak of the tip to the YSR state) produces minor peaks in the gap at $\pm (\Delta_\mathrm{T} - E_\mathrm{YSR})/e$.
Inelastic transitions (red dashed) to the YSR states produce pairs of peaks at $\pm (\hbar \omega + E_\mathrm{YSR} + \Delta_\mathrm{T})/e$.
Their relative intensities reflect those of the YSR resonances.
Their widths are much smaller than in (b) because they are not dictated by thermal broadening.} 
\label{prin}
\end{figure}

Here, we show that sharp resonances in the density of states of paramagnetic molecules on superconductors may be used to simultaneously increase the signal amplitude and the spectral resolution of inelastic tunneling spectroscopy.
The measurement concept may be introduced using energy diagrams of STM junctions (Fig.~\ref{prin}) between (a) two normal conductors and (b) a superconducting tip and sample.
In the former case, a vibrational mode of energy $\hbar\omega$ may be excited through inelastic tunneling transitions when $V>\hbar\omega$.
This leads to a minute stepwise change of the $I(V)$ slope that becomes more readily discernible in $d^2I/dV^2(V)$.
A model spectrum is displayed in Fig.~\ref{prin}(c).
It was calculated assuming a fairly large relative change of conductance ($\eta=10$~\%) due to the inelastic processes.
The peak width in $d^2I/dV^2$ is determined by the thermally broadened Fermi distributions so long as other instrumental effects can be neglected.
In the superconducting case [Fig.~\ref{prin}(c)], the local densities of states (LDOS) of tip and sample exhibit coherence peaks and gaps around the Fermi energies.
Moreover, sharp resonances within the superconducting gap of the sample may be induced by paramagnetic impurities. 
These Yu-Shiba-Rusinov (YSR) states are located at the energies $\pm E_\mathrm{YSR}$ around the sample Fermi energy $E_F = 0$ and strongly affect the current \cite{Yazd, Fran, nach21}. 
In the model calculation of Fig.~\ref{prin}(d), the YSR peak height exceeds the background (conductance $G_N$ outside the gap) by a factor of $\approx  20$. 
In some experiments, we achieved values of up to $\approx 50$.
When such a resonance is the final state of a transition the inelastic current is enhanced by the same ratio.
As a result, inelastic transitions prominently appear in differential conductance ($dI/dV$) spectra [(Fig.~\ref{prin}(d)] as asymmetric peaks that replicate the YSR resonances.
The relative change of $dI/dV$ due to a vibrational excitation is thus moved from the scale of a few percent \cite{stip, kath} to values up to $\approx 80$~\% %77
and the detection limit is improved by approximately one order of magnitude.

\begin{figure}[!ht]
\begin{center} \includegraphics[width=0.95\columnwidth]{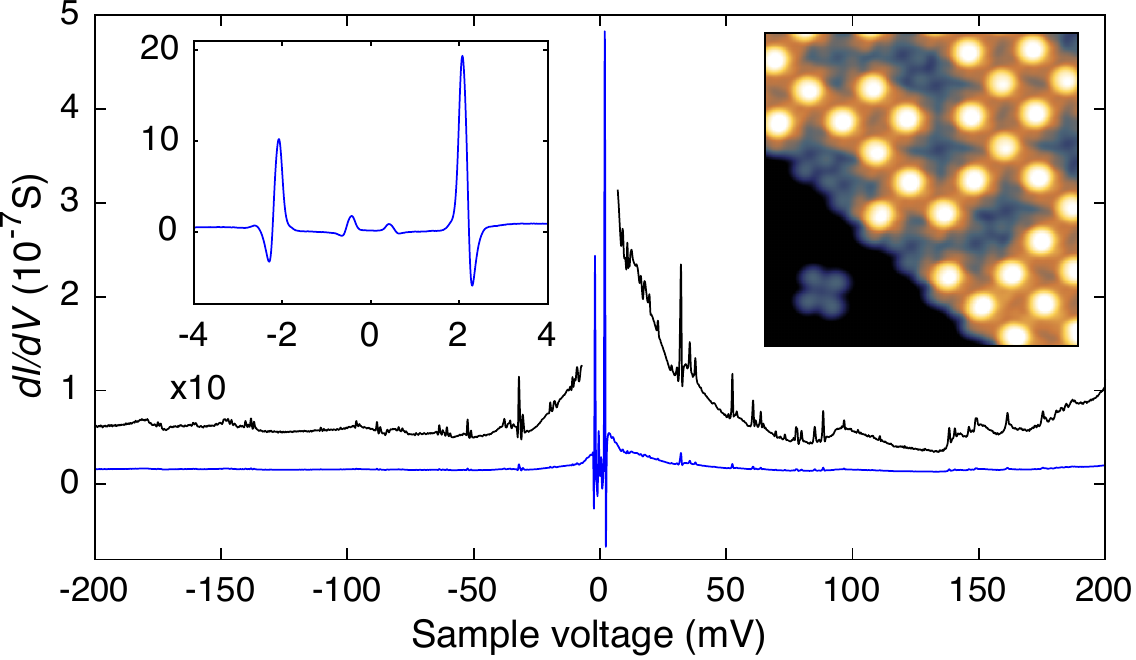} \end{center}
\caption{(color online) $dI/dV$ spectrum of PbPc on superconducting Pb(100).
The spectrum (lower blue line) was recorded above the Pb center of a molecule within a molecular island.
The most prominent spectral features are YSR resonances at $\pm 2$~mV (see also left inset).
A magnified view (multiplied by 10, black upper line) reveals a wealth of additional peaks that are due to vibration excitations.
The inset to the right shows a constant-current image of molecules assembled into an island with a square mesh.
Some molecules (blueish clover shaped patterns) lost their Pb ion.
The majority is intact with the Pb atom imaged as a bright protrusion.
The main spectrum is an average of 20 spectra recorded with an initial current of 4\,nA at $V=200$\,mV and a voltage modulation of 0.5\,mV$_{\mathrm{PP}}$.
The spectrum in the inset was measured with parameters 500\,pA at 5\,mV and 0.05\,mV$_{\mathrm{PP}}$ modulation.} 
\label{uber}
\end{figure}

We investigated Pb-phthalocyanine (PbPc) on superconducting Pb(100).
The experiments were performed in ultrahigh vacuum with scanning tunneling microscopes operated at 2.3 and 4.2\,K\@. 
The Pb(100) substrate was prepared in vacuo by repeated cycles of Ar$^+$ bombardment and annealing at about 530\,K\@.
PbPc molecules were sublimated from a Knudsen cell onto the Pb substrate held at room temperature.
The differential-conductance $dI/dV$ was measured using a lock-in amplifier with modulation voltages between 50 and 500\,$\mu$V$_{\mathrm{PP}}$ at a frequency of 831\,Hz.

Density functional theory (DFT) calculations were carried out for the free PbPc molecule as well as PbPc molecules adsorbed on Pb(100).
We also calculated the inelastic tunneling spectrum, $d^2I/dV^2$, using DFT combined with the non-equilibrium Green's function method with electron-phonon coupling taken into account using a current conserving lowest order expansion method.
All calculations were performed using QuantumATK \cite{quantumatk2021, Smidstrup_2019}.
In the Supplemental Material \cite{sm} we detail the calculation methods and results.

After deposition of PbPc at ambient temperature, the molecules arrange in a square mesh with two rotational domains and lie flat on the substrate as displayed in the right inset of Fig.~\ref{uber}.
The unit cell contains two molecules which are rotated by 8$^{\circ}$ (domain A) or 20$^{\circ}$ (domain B) with respect to each other.
Although some molecules (blueish clover shaped patterns) loose the Pb ion a majority stays intact with the ion either below or above the macrocycle.
 
The main experimental observation is the $dI/dV$ spectrum (blue lower line) recorded above a Pb center.
It appears flat over a wide voltage range and exhibits tall YSR resonances at $\pm 2$~mV (left inset).
Like in the case of H$_2$Pc \cite{ennea}, we find YSR states for arrays of PbPc but not for isolated molecules.
A magnified view (black upper line) reveals 46 clear features positioned symmetrically around zero bias.
They ride on a background that may be approximated by a superposition of a broad Lorentz-shaped resonance and the spectral signature of bulk phonons we observe on pristine Pb(100). 
Based on the analyzes presented below, we attribute the features to the excitation of molecular vibrations.
The YSR resonance amplitudes depend on the bias polarity and so do the related inelastic features.

\begin{figure}[!ht]
\begin{center} \includegraphics[width=0.95\columnwidth]{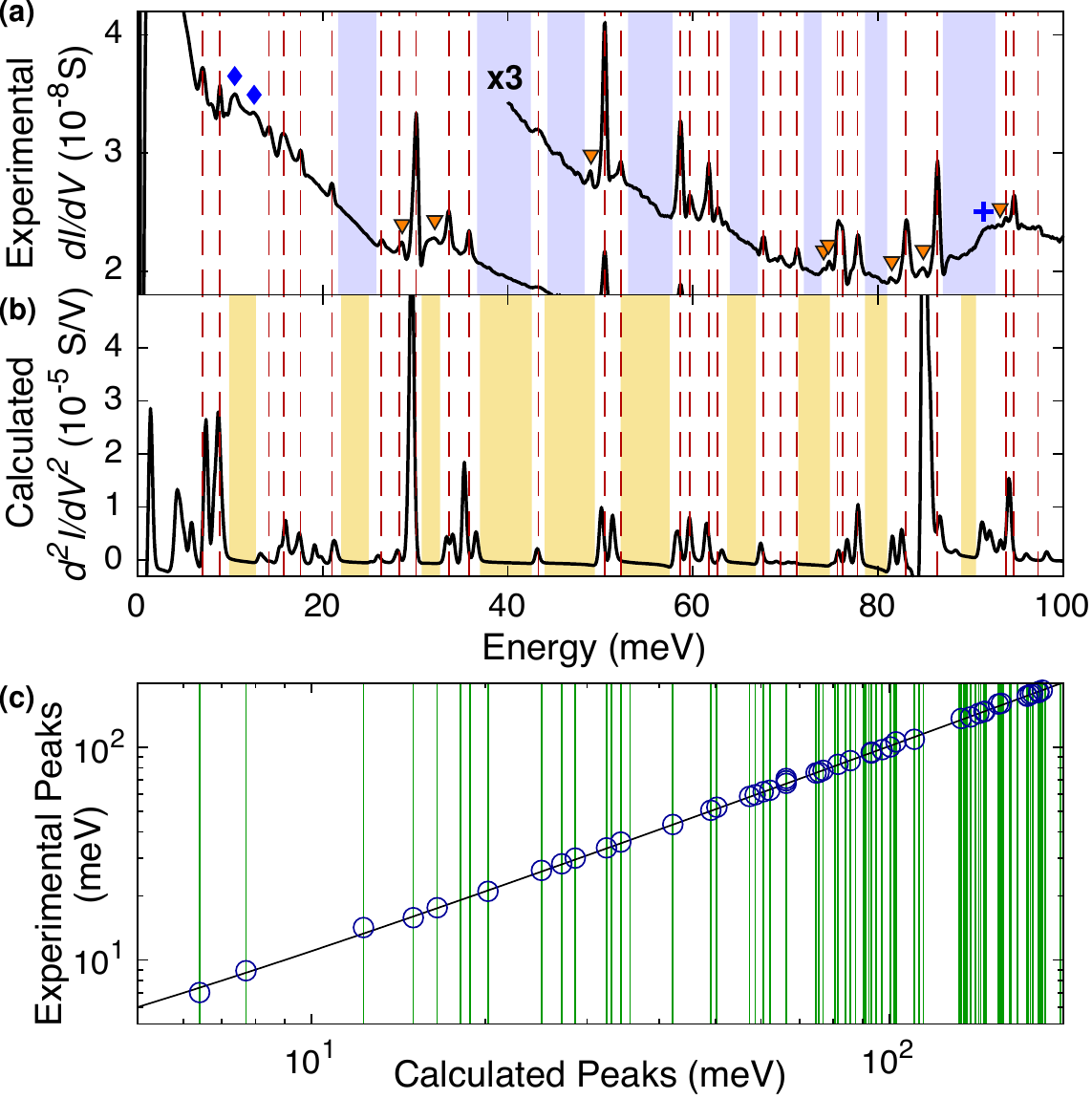} \end{center}
\caption{(color online) Comparison of experimental and calculated spectra.
(a) Low energy part of the $dI/dV$ data from Fig.~\ref{uber} shifted by the resonance position $E_\mathrm{YSR} + \Delta_\mathrm{T}$.
Vertical dashed lines indicate peaks.
Left of intense peaks, small copies are observed (orange triangles) due to thermally excited charge carriers in the tip.
In addition to molecular vibrations, bulk phonons, which are also visible also on pristine Pb(100), affect the spectrum at low energies.
Two broader peaks (blue diamonds) were omitted from the mode assignment for this reason.
A step at $\approx 90$\,meV (blue cross) coincides with a multi-peak structure in the calculated $dI^2/dV^2$.
(b) Calculated spectrum of $dI^2/dV^2$.
Except for a 1.0~meV shift, most of the calculated peak energies match the experimental ones thus enabling a tentative peak assignment.
Shaded areas indicate gaps in the vibrational spectra.
(c) Comparison of calculated peak energies (green vertical lines) with corresponding measured energies (vertical coordinate of blue circles).
Logarithmic axes are used to mimic constant relative uncertainties. 
The black line indicates the relation $E_{exp} = E_{calc}+1$~meV\@.
}
\label{vgl}
\end{figure}

Fig.~\ref{vgl} displays (a) the low energy part of the $dI/dV$ spectrum shifted by the energy $E_\mathrm{YSR}+\Delta_T$ ($2 \Delta_T$ is the excitation gap of the tip) and (b) a calculated spectrum of $dI^2/dV^2$. 
29 features are marked in the experimental data by vertical lines.
We find an excellent agreement with the calculated peak energies shifted by $+1.0$~meV\@.
This leads us to propose a tentative assignment of the experimental peaks to vibrational modes.
Data up to 200~meV are presented in Supplemental Fig.~S4 \cite{sm}.
As expected the differences between measurement and calculation are slightly larger at high energies.
Nevertheless, the similarity between the calculated and measured spectra remains as indicated in Fig.~\ref{vgl}(c).
All 46 experimental modes (blue circles) match with calculated ones (green lines).

The Supplemental Material \cite{sm} presents an overview and animations of the calculated vibrational modes of a PbPc monolayer on Pb(100). 
Our DFT calculations with different functionals showed that the vibrational energies are robust (Supplemental Fig.~S3 \cite{sm}).
The surface unit cell comprises two molecules supporting $2\times 171$ modes, many of them being almost degenerate pairs.
As expected for an array of interacting dipoles, the modes of the monolayer are blue-shifted compared to those of a single adsorbed molecule (Supplemental Fig.~S9 \cite{sm}), which in turn are red shifted from gasphase results because of the interaction with image charges \cite{masch, endl}.

In the calculated $d^2I/dV^2$ spectra we observe 74 peaks with intensities above an arbitrary threshold of 500~nA/V$^2$ corresponding to an effective cross section of $\eta \approx 0.07\%$.
Several of the corresponding modes (e.~g., 15, 63, 123, 205, and 239 at 6.3, 34.3, 83.9, 132.3, and 154.4~meV) exhibit the same B$_1$ symmetry as the two degenerate lowest unoccupied molecular orbitals (LUMOs), which are the most transmitting transport channels and, therefore, determine the electron transport \cite{paul}.
Most of the intense modes correspond to experimental peaks, but some calculated modes are close in energy and thus not individually resolved in $dI/dV$.
Presently, a more detailed analysis of the intensities is precluded by the fact that they drastically depend on the functional used and the resulting orbital energies.

At first glance it may seem surprising that electron transport through YSR states may be directly compared to transport channels involving the LUMOs.
However, spatial mapping of the YSR state shows detailed similarities with the LUMOs \cite{ennea}.
Still some characteristics of YSR states must be considered in analyzing the experimental data, as the interaction with the tip and an electric field can change their energy and line shape.
Our measurements on PbPc (Supplemental Fig.~S6 \cite{sm}) and published results show that the YSR states shift, broaden, and loose intensity when the tip is brought closer to the sample \cite{ysrofz, ysrofx}.
Taking this shift of the resonances into account we find distinct shifts of vibrational modes. 

\begin{figure}[!ht]
\begin{center} \includegraphics[width=0.95\columnwidth]{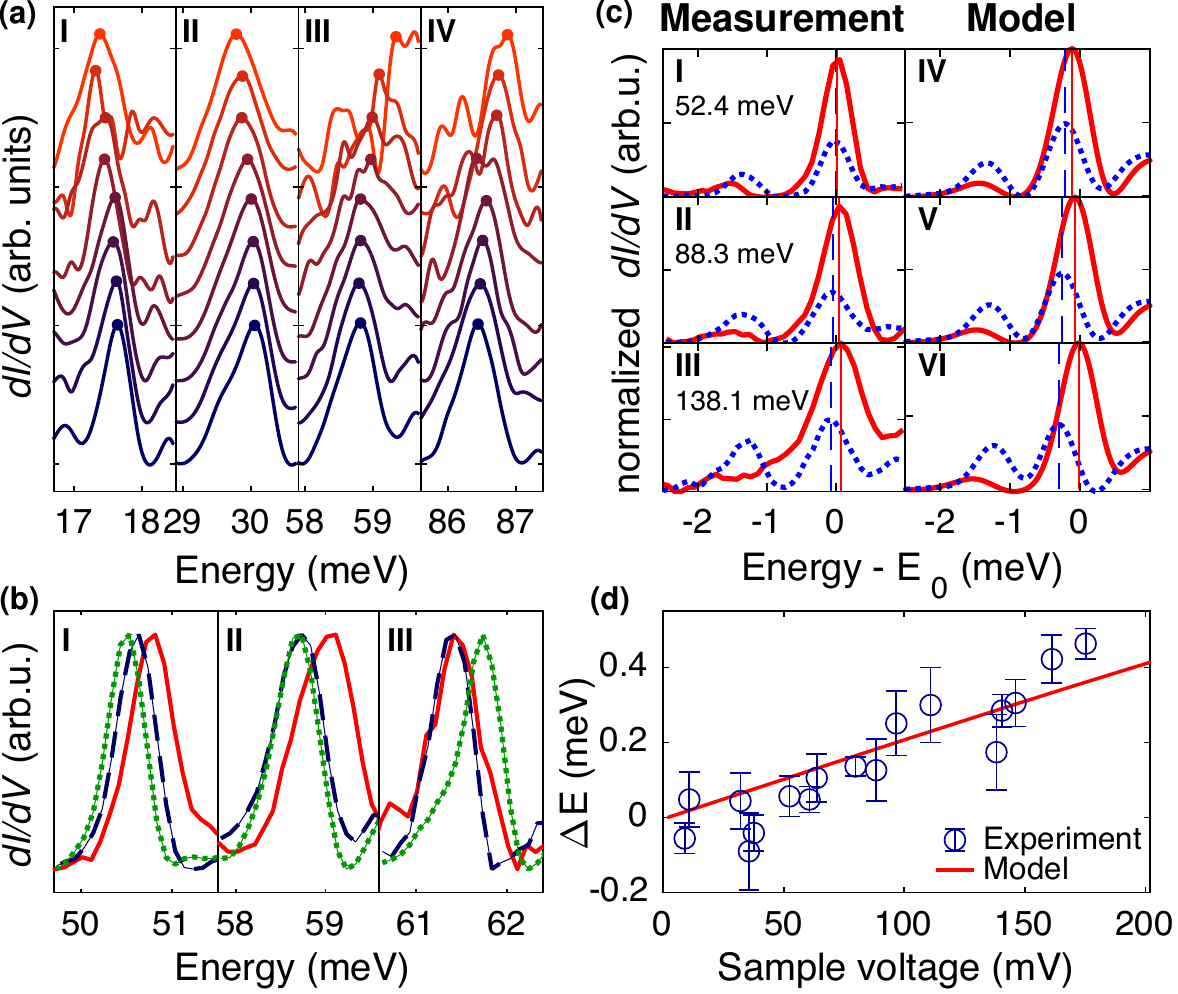} \end{center}
\caption{(color online) Spectral changes caused by tip, neighbors, and bias polarity.
(a) $dI/dV$ spectra of a PbPc molecule recorded over a range of tip heights.
For the bottom spectrum, the current feedback was disabled at $V=200$\,mV and $I=500$\,pA and the tip was brought 128\,pm closer to the sample.
From spectrum to spectrum the tip-sample separation was reduced by further 16~pm.
A sinusoidal modulation of 0.5\,mV$_{\mathrm{PP}}$ was used.
(b) $dI/dV$ spectra of PbPc molecules in different environments. 
The green dotted spectrum was recorded on a molecule with PbPc on nearest neighbor (NN) and next nearest neighbor (NNN) sites.
The blue dashed (red solid) curve shows data where Pb was absent from NN (NN and NNN) molecules.
(c) Left column: 
Experimental $dI/dV$ data of excitations at 52.5, 88.3 and 138.1~meV measured at positive (red solid) and negative (blue dotted) $V.$ $E_0 = E_\mathrm{YSR} + \Delta_T + \hbar\omega.$
The main peaks are due to tunneling from the occupied (unoccupied) coherence peak of the tip to the unoccupied (occupied) YSR state at $V>0$ ($V<0$).
Thermally excited charge carriers lead to a minor peak in each spectrum.
The positions of the main peaks (dashed vertical lines) do not exactly coincide and separate with $|V|$ in a linear fashion.
The relative heights of the major and minor peaks also change.
Right column: $dI/dV$ calculated with a model that considers a voltage dependence of the YSR states reproduces the experimental data.
(d) Polarity induced differences of peak positions evaluated from various excitations.
A model presented in the Supplemental Material reproduces the observed linear relation (Eq.~6).
Experimental parameters are identical to those in Figure~\protect{\ref{uber}}.} 
\label{appl}
\end{figure}

Fig.~\ref{appl} presents applications of the energy resolution of our method.
Panel (a) shows the evolution of mode energies with the tip-sample separation.
The modes near 18 and 30~meV exhibit red-shifts on the order of 0.1~meV while the modes near 58 and 86 meV undergo blue-shifts of similar magnitude.
The displacement patterns of the corresponding vibrational modes reveal some similarities.
The red-shifting modes (37/38, 55/56) involve mainly vertical atomic displacements while the blue-shifting modes (81/82, 123) are predominantly horizontal.
We tentatively attribute the blue shift to the attractive force exerted by the STM tip that increases the distance between the molecule and the substrate and thereby partially reverts the adsorption-induced red-shift of horizontal modes.
In addition, the tip softens the potential along the surface normal and thereby induces a red-shift of vertical modes.

Next, we reverse the roles of YSR states and vibrational excitations.
As the voltage between the tip and the sample may gate the molecule, the energy and peak height asymmetry of YSR states could be bias dependent.
In contrast to elastic tunneling processes, which probe YSR states at voltages close to zero, the vibration induced copies of the YSR resonances appear at higher voltages and therefore provide access to a possible gating effect.
Fig.~\ref{appl}(c) shows the spectra of three vibrational excitations at positive (red solid line) and negative (blue dotted line) voltages.
Each excitation gives rise to a pair of peaks at both polarities, the peak closer to zero bias being due to thermally excited electrons (holes) in the coherence peaks of the tip above (below) $E_F$.
Towards higher excitation energies the peaks separate in a linear fashion (Fig.~\ref{appl}(d), circles). 
The changes in peak positions and the shape of the inelastic signatures are well reproduced within a Bogoliubov-de Gennes model (Fig.~\ref{appl}(d), line).
To calculate inelastic tunneling spectra, the bias dependent density of states of the sample around $E_F$ is obtained using a Green's function approach that takes into account a constant potential scattering $W$ and an exchange coupling $J$, which depends linearly on the voltage.
The observed shift may be understood as follows.
At $V<0$, the LUMO is lowered relative to $E_F$ of the sample, its occupation increases, the YSR state moves deeper into the superconductor gap, and finally the peak height asymmetry is enhanced.

The interaction between a molecule and its local environment may also be addressed.
Figure~\ref{appl}(b) and Supplemental Figure S5 \cite{sm} show spectra of PbPc molecules surrounded by PbPc molecules on nearest neighbor (NN) and next nearest neighbor (NNN) sites (green dotted).
When the Pb ions are removed from either NN-molecules (blue dashed) or both, NN and NNN-molecules (red solid line), the vibrational modes observed on the center PbPc undergo small but clearly resolved shifts.
Three segments from spectra in Fig.~\ref{appl}(b) illustrate diverse evolutions.
Some modes shift continuously to lower (Fig.~\ref{appl}(b)\textsf{I}) or higher (Supplemental Fig.~S5 \cite{sm}) energies as the number of neighbors with Pb increases.
In other cases (Fig.~\ref{appl}(b)\textsf{II}), NNN PbPc induces a substantial red shift that is not further affected by NN Pb ions.
For the vibrations shown in Fig.~\ref{appl}(b)\textsf{III}, NNN PbPc molecules are required to produce a substantial blue-shift.

In summary, taking advantage of YSR states, the detection limit of vibrational spectroscopy with the STM is improved and the energy resolution is pushed beyond the thermal limit.  
Such resonances may be expected from molecules that carry a spin, be it due to intrinsic molecular properties or charge transfer involving the environment.
While modeling of the vibrational mode energies is fairly accurate, the inelastic intensities will require further attention.

\bibliographystyle{apsrev4-2}

\end{document}